\documentclass[10pt,conference,letterpaper]{IEEEtran}

\begin{document}

\title{Faster Transaction Commit even when Nodes Crash}

\author{\IEEEauthorblockN{Ryan Emerson}
\IEEEauthorblockA{School of Computing Science
\\Newcastle University, UK\\
Email: ryan.emerson@ncl.ac.uk} \and \IEEEauthorblockN{Paul
Ezhilchelvan} \IEEEauthorblockA{School of Computing Science
\\
Newcastle University, UK\\
Email: paul.ezhilchelvan@ncl.ac.uk}}\maketitle

\maketitle

\begin{abstract}
Atomic broadcasts play a central role in serialisable in-memory
transactions. Best performing ones block, when a node crashes, until
a new view is installed. We augment a new protocol for uninterrupted
progress in the interim period.
\end{abstract}

\section{Introduction}

This fast abstract is concerned with the sustainable performance of
1-copy serialisable transactions running on an 
in-memory database where data is partitioned and replicated over
the RAM of several nodes; no node holds all replicas of a given
partition nor the entire database (\emph{partial} replication).
In-memory databases are best suited to applications that require
frequent access to large data, mainly because they offer a superior
performance (e.g., by asynchronous disk writes) and can dynamically
scale (by changing the number of nodes whose RAM contributions
comprise the database infrastructure).

The combination of serialisability requirement and partial
replication constraint poses certain challenges that are effectively
addressed (see \cite{pstore}) by using atomic broadcasts, or
\emph{abcasts} for short. In brief, a transaction executes
read-optimistically, get its relative order through abcast, and
aborts only if its reads were made out-of-date by transactions
preceding it in the abcast order; however, if the transaction is read-only and its data is replicated locally, it is exempted from the out-of-date read scrutiny. Note that while this exemption is
certainly a performance enhancer, it ceases to exist whenever a
transaction execution spans over data items that are not stored locally as the data can be autonomously updated at other nodes.

Use of abcasts is shown \cite{InfiniTOM} to be highly effective
compared to the classical 2-Phase Commit approach. This comparative
study uses Red Hat's open-source in-memory database
\emph{infinispan} and considers all influential factors such as
abort rate, latency/throughput and the average number of nodes
involved in a transaction. The abcast protocol used in this study,
however, is chosen from a class of protocols, such as \cite{NewTop,
VSAbcast}, that work \emph{extremely} well in the absence of
crashes; when a node crashes, however, they block until a group
membership protocol delivers a new membership \emph{view} that must
also contain a \emph{virtually-synchronous} closure on the set of
messages that should be delivered in the old view \cite{VSAbcast}.
This can take in the order of seconds, e.g., JGroups uses a default
timeout of 10s to rule out false crash suspicions prior to
constructing the new view. Thus, the study, albeit comprehensive
otherwise, is valid only in the absence of node crashes.

The objective of our work, also sponsored by Red Hat, is to retain
the best crash-free performance and to mitigate the effects of
post-crash blocking. It is being accomplished in two stages: (i)
delivering abcast as a separate service rather than relying on each
transaction initiator itself to execute an abcast protocol with
nodes involved in its transaction ($\S$\ref{ABcastService}); and,
(ii) incorporating an \emph{insurance} abcast protocol that can be
effortlessly switched on or off whenever a crash is suspected or
whenever a new view is ready or the suspicion has turned out to be
false, respectively ($\S$\ref{BestEffort}). Before presenting design
challenges and the achieved/expected outcomes in each stage, we
highlight next the limited options available in achieving our
objective.

\section{Approach and Rationale}
\subsection{No Cheap Asynchronous ABcast Insurance}

Group-membership dependant (GMD for short) abcast protocols are
asynchronous: they do not assume bounds on message delays nor on
clock differences between nodes. When a node abcasts a message $m$,
recipients broadcast an \emph{ack} for $m$ promising that they will
only broadcast $m'$ with a time-stamp larger than the one $m.ts$
found in $m$. When an ack is received from \emph{every} member in
the group, $m$ is ready to be ordered as per $m.ts$. When all
members are operative, a GMD abcast can have the smallest latency of
one round-trip delay (when \emph{ack}s are broadcast instantly) and
the lowest message cost of 1 broadcast (when \emph{ack}s are
piggybacked).

The other class of asynchronous abcast protocols are quorum based
(QB for short). Crash tolerance is inherent in each design/execution
step: no effort is made to detect whether any node is \emph{truly}
crashed and measures are undertaken \emph{as though} at most less
than half the nodes can crash at any time soon.

Suppose that a GMD protocol is chosen as the normal abcast when no
crash is suspected and a QB one as the insurance whenever a crash is
suspected. Switch-over requires a virtually synchronous closure on
the normal stream of already ordered, and possibly delivered,
messages, i.e., constructing an agreed 'view' on stream closure for
operative nodes is essential for consistent switch over. So,
switch-over is computationally as 'heavy-weight' as in JGroups,
except for the long 10-second duration used there to ascertain an
actual crash. Using smaller timeouts can lead to false crash
suspicions, making switch-over unnecessary at times. No optimal
timeout exists to discern a slow node from a crashed one
\cite{Fischer1985}.

\subsection{Our Approach}
We use a proactive synchronous abcast protocol as the insurance.
Dedicated nodes implement abcast as a service to nodes executing
transactions. They keep their clocks synchronized within some known
accuracy $\epsilon$ with a high probability, using
\cite{ProbClockSync}. They timestamp each $m$ and $ack$ they
broadcast, which allows message delays to be pessimistically estimated. From the delays observed in the recent
past, each node $i$ estimates the worst case delay $d_i$ which it
encodes in its broadcast $m$.

When node $j$ has $m$ and is not aware of any other $m'$, $m'.ts
 \leq m.ts$, that is yet to be ordered, it orders $m$ (as per $m.ts$) by the GMD abcast rules
or after its clock time is $m.ts + D +\epsilon$, whichever is earlier; here, $D$ is some function of $d_i$ and other parameters
corresponding to various best effort protocol measures aimed at
making node $j$ be aware of any such $m'$. These measures are
outlined in Subsection \ref{BestEffort}.

Analysing the success of having a proactive synchronous protocol as
the insurance involves two cases: node $i$ is slow or crashed and
node $j$ has or knows of $m$ before $m.ts + D +\epsilon$ (Case 1) or
never knows of $m$ until $m.ts + D +\epsilon$ and orders another
$m'$, $m'.ts > m.ts$ (Case 2). Case 2 occurs when all best efforts
within the protocol are rendered ineffective by sharp increases in
communication delays. We seek to minimise the Case 2 probability to
as small as $10^{-6}$.

\section{Contributions: Completed and Expected}

\subsection{ABcast as a Service}\label{ABcastService}

When a node, \emph{Tx\_host} for short, that initiated a transaction
$Tx_i$ completes the execution, it sends a request to this
(external) ordering service for a global order to be put on $Tx_i$,
together with the list of all nodes participating in $Tx_i$. The
request is sent to one of the multiple dedicated servers
implementing the abcast service. The contacted server responds back
with a global order number for $Tx_i$ to \emph{Tx\_host} which, in
turn, forwards the order to all participating nodes.

A subtle issue here is to ensure that a participant node $i$ is not
forced to undergo a cascaded waiting when it is concurrently
participating in several transactions. Say, node $i$ participates
concurrently in $Tx_1$ and $Tx_2$ initiated by \emph{Tx\_host}$_1$
and \emph{Tx\_host}$_2$, respectively. Receiving just the order
number for one transaction, say, $Tx_1$ from \emph{Tx\_host}$_1$ but
not (yet) for $Tx_2$ does not allow node $i$ to determine if $Tx_1$
precedes $Tx_2$ or vise versa. So, the response of an order server
to a \emph{Tx\_host} includes, for each participant node listed in
the request, a short history of transactions preceding the one whose
ordering has been requested. So, when node $i$ receives the order
number and history for $Tx_1$, if $Tx_2$ is not in the history for
$Tx_1$, node $i$ can work on $Tx_1$, even if it has not yet received
the ordering details for $Tx_2$ from \emph{Tx\_host}$_2$.

The main advantages of abcast service are: \emph{Tx\_host} and
participant nodes are spared from executing an abcast protocol and
the protocol is not restricted to be leader-based; the main cost is:
time delay in contacting, and receiving the response from, the
service. We replicated the experiments of \cite{InfiniTOM} and the
results indicate using an external order service pays off when the
average number of nodes involved (\emph{Tx\_host} and participants)
exceeds  3.3. Thus, an external order service favours scalability
and protocol flexibility.

\subsection{Best Effort Design Aspects}\label{BestEffort}

The objectives are to (i) make a server node $i$ be aware of an
order request $m$ before its clock time $m.ts + D + \epsilon$ and
(ii) ensure that $D$ accommodates, as much as possible, delay
variations that might occur over and above the past estimate.

On the first objective, a 'broadcast' of $m$ consists of two
redundant broadcasts separated by some interval ($\eta$) and an
$ack$ incorporates the last sequence number of the broadcast
received from each server. The latter enables node $i$ to deduce any
broadcast it may be missing and postpone ordering of later messages
until 'gaps' are filled. The former enables a node to suspect that
all is not well with the first broadcast, if the second broadcast
has not been received within a certain timeout; it prompts a
proactive response by re-broadcasting the message on behalf of the
sender. (Care is taken to minimise proactive responses.) A recipient
server's response helps to complete a broadcast that may be rendered
partial due to sender crash and also to fill in the 'gaps'.

Value for $D$ is estimated as some function of $\eta$ and the
probability distribution of delays estimated in the past. The
function itself is designed to be pessimistic. Examples of pessimism
are: a broadcast is said to be complete when the second redundant
broadcast reaches recipients; the sender is \emph{always} assumed to
crash during the first redundant broadcast, leaving a recipient to
do both the redundant broadcasts on behalf of the sender. From the
cumulative distribution for $D$, we choose a value corresponding to
99.99\% probability.

With the ordering service implemented by 3 dedicated server nodes,
we observed no out-of-order failures for fairly-large request
arrival rates. However, when arrival rates increase beyond a
threshold, servers tend to saturate undermining our hypothesis that
future delay can be estimated reasonably safely based on the past
delay estimates. We are therefore currently implementing flow
control to avoid server saturation.

\bibliographystyle{IEEEtran}
\bibliography{HiTab}

\end{document}